\documentclass[prd,amsmath,amsfonts,showpacs,superscriptaddress,twocolumn,a4paper,floatfix]{revtex4}

\usepackage{amsmath}
\usepackage{amssymb}
\usepackage{latexsym}
\usepackage{dsfont}
\usepackage{calc}

\usepackage{subfigure}
\usepackage{amsmath,amssymb}
\usepackage{graphicx}
\usepackage{bm}
\usepackage{color}
\usepackage{epsfig}

\newcommand{\be}{\begin{equation}}
\newcommand{\ee}{\end{equation}}
\newcommand{\bs}{\begin{split}}
\newcommand{\es}{\end{split}}
\newcommand{\bi}{\begin{itemize}}
\newcommand{\ei}{\end{itemize}}
\newcommand{\bn}{\begin{enumerate}}
\newcommand{\en}{\end{enumerate}}
\newcommand{\bfi}{\begin{figure}}
\newcommand{\efi}{\end{figure}}
\newcommand{\komm}[1]{}
\newcommand{\tr}[1]{\textrm{tr}\,#1}
\newcommand{\hmu}{\hat{\mu}}

\newcommand{\veps}{\varepsilon}

\renewcommand{\Re}[1]{\textrm{Re}\,#1}

\begin{document}
\author{Jan M.\ Pawlowski} \affiliation{Institut f\"ur Theoretische
  Physik, Universit\"at Heidelberg, Philosophenweg~16, D-69120 Heidelberg, Germany}
\affiliation{Extreme Matter Institute EMMI, GSI Helmholtzzentrum f\"ur Schwerionenforschung GmbH, Planckstr.~1, D-64291
  Darmstadt, Germany} 

\author{Daniel Spielmann} \affiliation{Institut f\"ur Theoretische
  Physik, Universit\"at Heidelberg, Philosophenweg~16, D-69120 Heidelberg, Germany}
\affiliation{Extreme Matter Institute EMMI, GSI Helmholtzzentrum f\"ur Schwerionenforschung GmbH, Planckstr.~1, D-64291
  Darmstadt, Germany} 

\author{Ion-Olimpiu Stamatescu} \affiliation{Institut f\"ur Theoretische
  Physik, Universit\"at Heidelberg, Philosophenweg~16, D-69120 Heidelberg, Germany}
\affiliation{FEST, Schmeilweg~5, D-69118 Heidelberg, Germany}

\title{Lattice Landau gauge with stochastic quantisation}
\begin{abstract}
  We calculate Landau gauge ghost and gluon propagators in pure
  $SU(2)$ lattice gauge theory in two, three and four dimensions. The gauge fixing method we use,
  sc.\ stochastic quantisation, serves as a viable alternative approach
  to standard gauge fixing algorithms. We also investigate the
  spectrum of the Faddeev-Popov operator. At insufficiently accurate
  gauge fixing, we find evidence that stochastic quantisation samples
  configurations close to the Gribov horizon. Standard gauge fixing
  does so only at specific parameters; otherwise, there is a clear
  difference. However, this difference disappears if the gauge is
  fixed to sufficient accuracy. In this case, we confirm previous lattice
  results for the gluon and ghost propagator in two, three and four dimensions.
\end{abstract}
\maketitle

\section{Introduction}\label{sec:intro}
Confinement is one of the key properties of QCD, but its explanation
from first principles still remains to be found. While a linearly
rising heavy quark-antiquark potential is readily simulated on the
lattice, evidence for or against the candidates for the underlying
confinement mechanism is more elusive. Some prominent confinement
scenarios involve topological defects, like magnetic monopoles or
center vortices, and are best described in specific gauges amiable
towards the definition of the related defects. It is this gauge
dependence which makes it so difficult to pin down the confinement
mechanism. There is some evidence that such scenarios are compatible
with approaches like the Gribov-Zwanziger scenario
\cite{Gribov:1977wm,Zwanziger:1994yu,Zwanziger:2003cf}, according to
which configurations in the vicinity of the Gribov horizon account for
confinement. This scenario makes specific predictions for QCD's
Green's functions, e.g.\ in Landau gauge that the gluon propagator
vanishes in the infrared (IR), the horizon condition. In turn the ghost dressing
function is IR divergent, called ghost enhancement. The Kugo-Ojima
scenario \cite{Kugo:1979gm} has similar implications. It aims at
describing confinement by the BRST quartet mechanism and requires
well-defined global BRST charges. Both scenarios put restrictions
on the infrared behaviour of ghost and gluon propagators if
confinement is present in the theory. In turn it has been shown
recently that ghost and gluon propagators with a sufficiently large
IR suppression of the gluon and an IR enhancement of the
ghost lead to quark confinement indicated by center symmetry 
\cite{Braun:2007bx}. The related IR suppression of the gluon
propagator has been seen in both, lattice studies and continuum
approaches, see e.g.\ \cite{Fischer:2008uz,vonSmekal:2008ws}.

The quantities relevant for such confinement scenarios, like the
aforementioned Green's functions, depend on the chosen gauge. Hence
these investigations are complicated by the Gribov problem. This problem relates
to the fact that local gauge conditions do not single out a unique
gauge copy \cite{Gribov:1977wm,Singer:1978dk}, but allow for an
abundance of Gribov copies satisfying the gauge condition on the same
gauge orbit.

Interestingly, Landau gauge seems not to fix uniquely the infrared
behaviour of the propagators. In continuum computations with
functional methods such as Dyson-Schwinger equations (DSE) and
functional RG flows one observes a one-parameter family of
solutions. The free parameter can be linked to the zero momentum value
of the gluon propagator in the infinite volume limit, see
\cite{Fischer:2008uz}. It is suggestive to relate this parameter to
global properties of the gauge fixing as Landau gauge still has
remnant gauge degrees of freedoms. In the continuum, this infrared
boundary condition is implemented by specifying a non-perturbative
renormalisation condition, see \cite{Fischer:2008uz}. It has been
demonstrated recently that a corresponding freedom exists also on a
finite lattice, and that it is directly connected to the
non-perturbative completion of the Landau gauge condition,
\cite{Maas:2009se}. Indications for this freedom have already been
seen in the strong coupling limit
\cite{Sternbeck:2008mv,Sternbeck:2008na,Sternbeck:2008wg} where the
effect is rather dramatic \cite{mpssvs}. This high sensitivity of the
results to the global details of the gauge fixing, in particular for
the ghost propagator, calls for a systematic evaluation of the details
of the thermodynamical limit of Landau gauge fixings on the
lattice. In particular the question of whether the Gribov-Zwanziger
scenario is indeed implemented within specific gauge fixings is of
importance.

In the present work, we compute ghost and gluon propagators in Landau
gauge Yang-Mills theory on the lattice within stochastic
quantisation \cite{Parisi:1980ys,Batrouni:1985jn,Damgaard:1987rr}, an alternative to standard Monte Carlo simulations. It has recently also received renewed interest e.g.\ for simulations at finite density \cite{Aarts:2008rr,Aarts:2008wh} or in real time \cite{Berges:2005yt}. Here, we employ stochastic quantisation for gauge fixing \cite{Zwanziger:1981kg,Seiler:1983rc,Seiler:1984pp,Rossi:1987hv,Nakamura93,Mizutani:1994za,Aiso:1997au,Nakamura:2002ki,Nakamura:2003pu}. In such an approach, a gauge fixing force drives the
Langevin evolution towards the first Gribov region. This potentially
opens new possibilities to circumvent the Gribov problem. It also
furthers the insights and allows for a better understanding of the
underlying structures. At insufficiently accurate gauge fixing, we
find evidence that stochastic quantisation samples configurations
close to the Gribov horizon. Standard gauge fixing does so only for
specific parameter choices; otherwise, there is a clear
difference. While this difference disappears if the gauge is fixed to
sufficient accuracy, the sampling close to the Gribov horizon makes it
unlikely that stochastic gauge fixing systematically suppresses
contributions from the Gribov horizon. Such a suppression would have
the potential of invalidating the argument that in the thermodynamical
limit only configurations at the Gribov horizon matter. In summary,
our analysis provides strong evidence against one of the reasons for a
failure of lattice gauge fixings to reproduce the Gribov-Zwanziger
scenario.

In the second section we define ghost and gluon propagators and
discuss their general scaling properties. In the third section we
introduce our implementation for stochastic quantisation and
stochastic gauge fixing. Some of the intricacies of such an approach
are then illustrated within a simple finite-dimensional model in the
fourth section. In the fifth section we present and discuss our
results on ghost and gluon propagators in two, three and four
dimensions, which we supplement in the subsequent section with some
results on the spectrum of the Faddeev-Popov operator. We close with a
short summary and discussion of our results.

\section{Infrared behaviour of the ghost and gluon
  propagator}\label{sec:irprops}
We are interested in the infrared behaviour of Green's functions in Landau gauge Yang-Mills theory. Landau gauge is defined via the condition,
\begin{equation}
  \partial_\mu A_\mu^a=0,
\end{equation}
which is subject to the Gribov ambiguity. In Landau gauge, the gluon
and ghost propagator have the form
\begin{eqnarray}
  (D_\text{gl})^{ab}_{\mu\nu}&=&\delta^{ab}\left(\delta_{\mu\nu}-
    \frac{q_\mu q_\nu}{q^2}\right)D_\text{gl}(q^2),\\
  (D_\text{gh})^{ab}&=&-\delta^{ab}D_\text{gh}(q^2), 
\end{eqnarray}
they are parametrised by the two scalar functions
$D_\text{gl/gh}(q^2)$. The standard dressing functions are given by $q^2 \,D_\text{gl/gh}(q^2)$. Functional methods in the continuum such as DSE
\cite{vonSmekal:1997is,vonSmekal:1997vx,Lerche:2002ep,%
  Alkofer:2004it,Zwanziger:2009je,Huber:2009tx,Alkofer:2000wg,Fischer:2006ub}, stochastic quantisation \cite{Zwanziger:2001kw,Zwanziger:2002ia,Zwanziger:2003cf} and the functional renormalisation group
\cite{Pawlowski:2003hq,Fischer:2004uk,Fischer:2008uz,Litim:1998nf,Pawlowski:2005xe,Gies:2006wv} allow for a
scaling solution which is unique \cite{Fischer:2006vf,Fischer:2009tn,Alkofer:2008jy} and in line
with global BRST-symmetry
\cite{Fischer:2008uz,vonSmekal:2008ws,PawlowskiSmekal}. As mentioned
in the introduction, there are strong arguments for a one-parameter
family of decoupling solutions where scaling is violated, see \cite{Boucaud:2006if,Boucaud:2008ji,Boucaud:2008ky,Dudal:2007cw,%
  Dudal:2008sp,Sorella:2009vt,Dudal:2009bf,Aguilar:2004sw,Aguilar:2007nf,Binosi:2007pi,Aguilar:2008xm,Cornwall:1981zr,Fischer:2008uz,%
  Fischer:2008yv,Kondo:2009ug,Kondo:2009gc,Kondo:2009wk,Kondo:2009qp,Maas:2009se} for related work. The unique scaling solution is
indeed the scaling endpoint of this one-parameter family, and the
corresponding propagators obey power laws in the infrared,
\begin{equation}
\lim_{q^2\to 0}D_\text{gl/gh}(q^2)\sim\frac{1}{(q^2)^{\kappa_{A/C}+1}}.
\end{equation}
In $d=4$ dimensions, the exponents satisfy the scaling relation
\begin{equation}\label{kAkC4D}
\kappa_A=-2\kappa_C.
\end{equation}
It stems from non-renormalisation of the ghost-gluon-vertex
\cite{Taylor:1971ff}, which has been confirmed on the lattice
\cite{Cucchieri:2004sq,Bloch:2003sk} as well as from DSE
\cite{Schleifenbaum:2004id}. When varying $d$, \eqref{kAkC4D}
generalises to
\begin{equation}\label{kAkCvarD}
\kappa_A=-2\kappa_C+(d-4)/2.
\end{equation}
Assuming a bare ghost-gluon vertex, the predicted values for $\kappa_C$, which is also referred to simply as $\kappa$,
are $\kappa=\frac{1}{98}\left(93-\sqrt{1201}\right)\approx 0.6$ in
$d=4$, $\kappa\approx 0.4$ in $d=3$ and $\kappa=0.2$ in $d=2$
\cite{Lerche:2002ep}.\par

The aforementioned decoupling solutions imply that both the gluon
propagator and the ghost dressing function are finite in the
infrared. In contrast to the scaling solution, the decoupling solutions 
seem to be incompatible with global BRST invariance 
\cite{Fischer:2008uz}, see
\cite{vonSmekal:2007ns,vonSmekal:2008en,vonSmekal:2008ws} for progress
towards a lattice BRST formulation.  Nevertheless, both solutions exhibit confinement, as both satisfy the confinement criterion
put forward in \cite{Braun:2007bx} and both lead to positivity
violation for the gluon. Positivity violation has been explicitly
confirmed on the lattice in $d=3$ \cite{Cucchieri:2004mf} and $d=4$
\cite{Sternbeck:2006cg,Bowman:2007du} and also from DSE
\cite{Alkofer:2003jj}.

\subsection{Lattice simulations in $d=2,3,4$}
Lattice simulations have so far found no clear evidence of a scaling
behaviour, with the possible exception of the two-dimensional case
\cite{Maas:2007uv,Cucchieri:2007rg}, which lacks dynamics
\cite{'tHooft:1974hx}. Very recently it has been shown that the
scaling behaviour might also be absent in two dimensions for the
lattice Landau gauge fixings used so far \cite{mpssvs}. In four
dimensions, lattice simulations of $SU(2)$ or $SU(3)$
\cite{Sternbeck:2005tk,Ilgenfritz:2006he,Sternbeck:2006cg,%
  Cucchieri:2007md,Bogolubsky:2007ud,Cucchieri:2007rg,%
  Bogolubsky:2008mh,Bogolubsky:2009dc} yield an IR finite gluon
propagator even on volumes up to $(27\,\text{fm})^4$. DSE studies on a
torus suggest that this volume could suffice for the correct infinite
volume behaviour to be visible \cite{Fischer:2007pf}. While dissenting
results concerning IR finiteness exist
\cite{Oliveira:2007dy,Oliveira:2007px,Oliveira:2008uf,Oliveira:2009eh},
they tend to yield an IR exponent closer to $\kappa=0.5$, implying an IR finite propagator, than to the predicted value
of $\approx 0.6$.\par Also for $d=3$, lattice results clearly favour
decoupling, e.g.\ \cite{Cucchieri:2003di}, even at $(85\,\text{fm})^3$
\cite{Cucchieri:2007md}. Recently, following a discussion of
non-perturbative ambiguities of the Landau gauge condition
\cite{Maas:2008ri}, evidence has been found that the IR behaviour
especially of the ghost propagator might strongly depend on the choice
of the Gribov copy on the residual gauge orbit
\cite{Maas:2009se}. This corresponds to the choice of
boundary conditions in functional methods \cite{Fischer:2008uz}.\par
In the strong coupling limit
\cite{Sternbeck:2008mv,Sternbeck:2008na,Sternbeck:2008wg,Cucchieri:1997fy}, simulations have likewise also been performed for $d<4$ \cite{Cucchieri:2009zt,mpssvs}. The interpretation of the results is still under debate. It
has been argued that the lattice data are compatible with a scaling
branch for a subset of lattice momenta \cite{Sternbeck:2008mv}, but
also that they definitely disprove such a behaviour
\cite{Cucchieri:2009zt}. Very recent additional results provide strong
evidence for the severity of the Gribov problem, especially regarding
the ghost propagator \cite{mpssvs}.

\subsection{Gauge fixing problem on the lattice}
Since the ghost and gluon propagator are gauge-dependent quantities,
their calculation requires to sample configurations from the gauge
fixing surface $\Gamma=\{A_\mu(x)|\partial_\mu A_\mu(x)=0\}$. Fixing
to lattice Landau gauge amounts to maximising a gauge fixing
functional whose Hessian is the Faddeev-Popov operator $-\partial_\mu
D_\mu^{ab}[A]$, where $D_\mu$ is the covariant derivative. Thus, after
gauge fixing, all configurations are located inside the Gribov region
$\Omega=\{A_\mu(x)|-\partial D[A]>0\}$. However, $\Omega$ still
contains gauge copies, in contrast to the fundamental modular region
$\Lambda\subset\Omega$. On the lattice, reaching $\Lambda$ is equivalent to obtaining the \emph{global} maximum of the gauge fixing
functional
\begin{equation}\label{stdgff}
R=\frac{1}{N_cVd}\sum_{x,\mu}\Re{\tr{U_\mu(x)}}.
\end{equation}
This poses an NP-hard optimisation problem, equivalent to finding the
ground state of a spin glass
\cite{Marinari:1991zv,Barahona82}.

\section{Stochastic quantisation on the lattice}\label{SQL}
\subsection{Stochastic gauge fixing}

As already discussed in the introduction, there is quite some evidence
that Landau gauge fixing on the lattice still severely suffers from
the Gribov problem. This already motivates the choice of an
alternative lattice gauge fixing algorithm.\par Stochastic quantisation is
a method introduced in a more general context in
\cite{Parisi:1980ys}. Within stochastic quantisation gauge fixing can be implemented via a
gauge fixing force \cite{Zwanziger:1981kg}. An infinitesimal gauge
transformation in the Langevin equation for the gauge field is given
by
\begin{equation}
  \frac{\partial A^a_\mu}{\partial t}=-\frac{\delta S_\text{YM}}{\delta A^a_\mu}
  +D^{ab}_\mu v^b+\eta_\mu^a.
\end{equation}
Here, $v^b=\alpha^{-1}\partial_\mu A_\mu$ with a gauge fixing
parameter $\alpha$, and $\eta_\mu^a$ is Gaussian white noise with the
properties
\begin{equation}\begin{split}
    \left\langle\eta_\mu^a(x,t)\right\rangle&=0,\\\left\langle
      \eta_\mu^a(x,t)\eta_\nu^b(x',t')\right\rangle&=2\delta_{ab}\delta_{\mu\nu}
    \delta(t-t')\delta(x-x').
\end{split}\end{equation}
Due to Zwanziger's gauge fixing term, the equilibrium configurations
of this equation are fixed to Landau gauge. It can be shown
\cite{Mizutani:1994za} and has been demonstrated
numerically \cite{Seiler:1983rc} that this equilibrium is stable only
inside $\Omega$.\par 

While stochastic gauge fixing cannot be expected to restore global
BRST invariance, it is a priori possible that it might sample the
configuration space differently from standard gauge fixing. E.g.,
it might give us a {\it uniform} gauge fixing on the first Gribov
region $\Omega$. Note in this context that the statistical arguments
in favour of the Gribov-Zwanziger scenario only work if the gauge
fixing does not single out some region inside the first Gribov region
in the thermodynamical limit. A uniform distribution avoids such a
possibility.\par
On the other hand, stochastic quantisation is unlikely to provide us
with a tool for a unique gauge fixing in the fundamental modular
region $\Lambda$, since it does not even aim at the global
maximisation of the gauge fixing functional $R$. But this fact might
not harm its prospects for two reasons: First, even sophisticated
algorithms designed to better approximate the global maximum of the
functional $R$ do not yield the scaling solution on the lattice
\cite{Bogolubsky:2007bw,Bogolubsky:2007pq,Bornyakov:2008yx,Bogolubsky:2008mh,%
  Bogolubsky:2009dc}. Second, the FMR and the Gribov region should be
equivalent in the thermodynamical limit according to a conjecture by
Zwanziger \cite{Zwanziger:2003cf}. So, if stochastic quantisation
leads to a uniformly covered Gribov region, this might be already
sufficient.

\subsection{Lattice formulation}
We simulate quenched $SU(2)$ with the group elements conveniently
parametrised as
\begin{equation}
U_\mu(x)=u_\mu^0(x)+i\sigma^a u_\mu^a(x).
\end{equation}
Unquenching effects are discussed in \cite{Ilgenfritz:2006he}, and
$SU(3)$ is treated in
\cite{Cucchieri:2007zm,Oliveira:2007px,Sternbeck:2007ug,%
  Bogolubsky:2007ud}.  A comparison of $SU(3)$ and $SU(2)$ is
specifically interesting as most continuum implementations with
functional methods use approximations which are insensitive to
$1/N_c$-effects. \par
The lattice formulation of the Langevin equation, initially without
Zwanziger's drift term, is \cite{Batrouni:1985jn}
\begin{equation}\label{localupdate}
  U_\mu(x)\to\exp\left(i\sigma^aR^a_{\mu x}\right)U_\mu(x),
\end{equation}
with $R$ composed of the Lie derivative of the standard Wilson
plaquette action \cite{Wilson:1974sk}, providing the dynamical drift force $F$, and a stochastic term,
\begin{eqnarray}
  R^a_{\mu x} &=& \veps F^a_{\mu x}+\sqrt{\veps}\eta^a_{\mu x}\\
  F^a_{\mu x} &=& i\nabla^a_{\mu x}S[U].
\end{eqnarray}
In addition, we have implemented a random walk method, whose results
agree with those obtained from Langevin evolution. In this case, the
form of an individual update is
\begin{equation}
  U_\mu(x)\to\exp(i\sigma^aA_\mu^a(x))U_\mu(x),
\end{equation}
and $A_\mu^a(x)$ is determined, starting from zero, by accepting
updates
\begin{equation}
  A_\mu^a(x)\to A_\mu^a(x)\pm\eta
\end{equation}
with probability
\begin{equation}
  p=\frac{1}{2}\left(1\pm\tanh\left(\frac{1}{2}\eta\, F^a_{\mu x}\right)\right).
\end{equation}
Regardless of the method used for the dynamic updates, Zwanziger's additional 
drift force is implemented as a lattice gauge transformation,
sc.\ $U_\mu(x)\to\Omega(x)U_\mu(x)\Omega^\dagger(x+\hmu)$ with
\cite{Rossi:1987hv,Nakamura93,Mizutani:1994za}
\begin{equation}
  \Omega(x)=\exp\left(-i\frac{\beta}{2N_c}
      \Delta^a\sigma^a\frac{\veps}{\alpha}\right).
\end{equation}
Here, $\Delta^a$ is the lattice version of the Landau gauge condition,
\begin{equation}
\Delta^a(x)=\sum_{\mu=0}^{d-1}(u_\mu^a(x)-u_\mu^a(x-\hmu)).
\end{equation}
By means of this implementation, the gauge force moves are not only
tangential to a gauge orbit, but strictly on the orbit.\par

We interchange the updates \eqref{localupdate} and the gauge
transformations `locally', i.e., such that a gauge transformation of
all affected link variables is performed after each single link
update. This serves to minimize the distance to the gauge fixing
surface during the updates, measured by $\Delta^2$, the average of
$\Delta^2(x)$. However, $\Delta^2$ does not decrease monotonically
when $\alpha$ is lowered (fig. \ref{medDS_vsa_forpaper}), which
presumably is due to an `overshooting' effect.\par
For a high precision of gauge fixing, we usually amend an interchange
of dynamical and gauge fixing steps as described above by standard
stochastic overrelaxation (STOR) \cite{ForcrandGupta89}. While this
goes, strictly speaking, beyond the scheme of stochastic gauge fixing,
it hardly affects the `locality' of the latter. An
alternative possibility would be to choose a step size small enough to
render the `amendment' dispensable, cp. the decrease of $\Delta^2$
with the step size in fig. \ref{medDS_vsa_forpaper}. But this would require so small a step size that the autocorrelation time would become unacceptably large for practical simulations.\par In contrast,
we refer by the term `standard gauge fixing' to a usual heat-bath
thermalisation \cite{Creutz:1980zw} followed by STOR. This is a
`global' procedure, as the thermalisation takes place far away from
$\Gamma$, i.e., at large $\Delta^2$. The difference is sketched in
fig. \ref{SQvsTGF_smaller}.

\bfi
\centering
\includegraphics[clip,width=0.92\columnwidth]{figs/medDS_vsa_forpaper.eps} 
\caption{Illustration of the change of $\Delta^2$ with the gauge
  fixing parameter $\alpha$ on a $48^2$ lattice for two different
  random walk step sizes.}
\label{medDS_vsa_forpaper}
\efi
\bfi
\centering
  \includegraphics[clip,width=0.92\columnwidth]{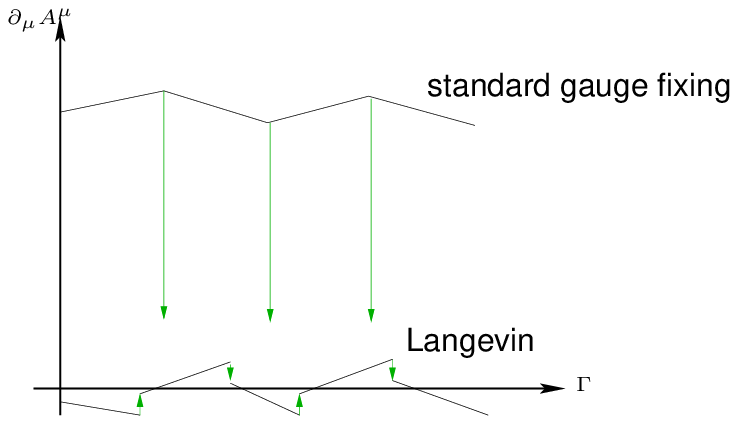}
  \caption{Qualitative illustration of a difference between stochastic
    and standard gauge fixing. The green arrows denote gauge fixing
    steps, the black lines connecting them denote dynamic updates. A
    similar picture appeared already in \cite{Nakamura:2003pu}.}
\label{SQvsTGF_smaller}
\efi

\subsection{Return cycles}\label{subsec:retcyc}
\begin{figure*}
  \centering \mbox{\includegraphics[clip,width=0.92\columnwidth]{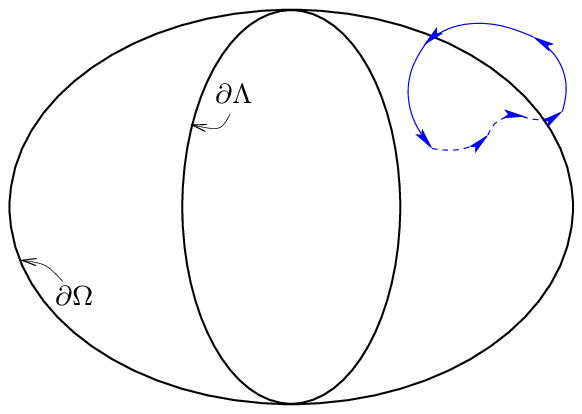}
    \qquad\qquad\includegraphics[clip,width=0.92\columnwidth]{figs/DSetLam_ty18_ls24_2d_ab_forpaper.eps}}
\caption{\emph{Left (a):} Qualitative illustration of a return
  cycle. \emph{Right (b):} Numerical data. Approach to $\Omega$ during a
  return cycle with the Langevin algorithm ($24^2$). Black: Lowest
  nontrivial FPO eigenvalue ($\hat{=}$ distance to $\partial\Omega$),
  red: distance to gauge fixing surface. }
\label{fig:retcy}
\end{figure*}

It has early been demonstrated \cite{Seiler:1983rc} that a start
outside the Gribov horizon leads to `return cycles' during which the
distance from the gauge fixing surface temporarily increases, i.e.,
the system returns to the Gribov region by a path outside the gauge
fixing surface. This is expected since only gauge-fixed configurations
inside the Gribov region are stable equilibrium configurations. This
is by itself not a special property of the gauge force, as it is
shared by every algorithm that performs numerical gauge fixing. But a
priori, the return cycles might play a role in bringing about a
certain distribution in configuration space, see
subsec. \ref{subsec:FPOS_results}.\par
The existence of return cycles is readily confirmed by the
numerics. By starting the system outside the Gribov region, e.g., at
sufficiently large constant gauge field, and measuring the lowest nontrivial 
eigenvalue of the Faddeev-Popov operator during the stochastic
evolution, we can indeed confirm that the configurations move towards
the Gribov region, in accordance with an intuitive picture, see
fig. \ref{fig:retcy}. The lowest eigenvalue $\lambda_0$ increases
monotonically, while the distance from the gauge fixing surface
reaches a maximum before decreasing again, as indicated by the
behaviour of
$\Delta^2$.\par

\section{Stochastic quantisation in a toy model}
Before we present the bulk of our results on stochastic gauge fixing
in $SU(2)$ lattice gauge theory in sec.~\ref{sec:results}, we find it
useful to illustrate some issues related to stochastic gauge fixing and the
configuration space in QCD by means of a simple toy model.
\subsection{Motivation}
The toy model we employ is a slightly modified generalisation of a
two-dimensional model used to illustrate some properties of stochastic
gauge fixing in \cite{Seiler:1984pp}. Note that a different toy model for stochastic gauge fixing has been studied in \cite{Langfeld:2003zia}. As a modification of the model from \cite{Seiler:1984pp}, we have
chosen the sign of the action such that the dynamical effects and the
gauge fixing effects can be clearly distinguished. The generalisation
consists in extending the model to a higher number of dimensions. It
is motivated by the possibility of observing an `entropy effect' that
leads to an accumulation of configurations close to the `Gribov
horizon' of this model, since such configurations should account for
confinement in QCD according to the Gribov-Zwanziger scenario.  Hence
the model could give an idea about the dependence on the three effects:
the gauge fixing force, the dynamics of the action and the entropy. 
\subsection{The toy model}
In the model, the variables $x_i\in\mathds{R}$ ($i\in\{1,\ldots,n\}$)
and $y\in\mathds{R}$ are subject to two forces, a gauge fixing force
and a force derived from an action. The former reads
\begin{equation}\label{gfforce_highdim}
  \begin{pmatrix}\dot{y}\\\dot{x_i}\end{pmatrix}=
  \begin{pmatrix}K_0^g\\K_i^g\end{pmatrix}=-\frac{1}{\alpha}
  \begin{pmatrix}(1-x^2)y\\2x_iy^2\end{pmatrix},
\end{equation}
where $x^2=\sum_{i=1}^n x_i^2$ and the gauge fixing parameter $\alpha$
is positive. Notice that this force is not the gradient of an
action. It drives the configurations towards the gauge fixing surface
$y=0$. The set
\begin{equation}\label{GribovRegion}
  \Omega=\left\{\left.\begin{pmatrix}y\\
        \vec{x}\end{pmatrix}\right|y=0,x<1\right\}
\end{equation}
may be referred to as the `Gribov region' of this model. This
terminology is sensible because the Gribov region is where all
configurations are in the limit of perfect gauge fixing -- and because
the asymptotic distribution for $\alpha\to 0$ is
\begin{equation}\label{asymp_alpha_inf}
  dP(x,y)\propto x^{n-1}(1-x^2)\theta(1-x^2)\delta(y)dxd\Omega dy,
\end{equation}
where $d\Omega$ is understood as the differential solid angle in $n$
dimensions.\par
The `Faddeev-Popov determinant' of this model is
$-\alpha^{-1}(1-x^2)$, which is proportional to
\eqref{asymp_alpha_inf} only for $n=1$ and vanishes, like
\eqref{asymp_alpha_inf}, linearly near the horizon $x=1$. The FP
operator is obtained from the linearisation $\frac{d}{dt}v=hv$, with
$v=0$ the gauge condition, thus here $v=y$.\par
The distribution \eqref{asymp_alpha_inf} is maximal at
$x=\sqrt{\frac{n-1}{n+1}}$. The maximum approaches the `Gribov
horizon' $\partial\Omega$ as $n$ grows. It is interesting to see
whether stochastic gauge fixing supports this effect. Obviously, the
gauge fixing force is orthogonal to the gauge fixing surface in its
immediate vicinity, see fig. \ref{forcefield_gauge_2a15_2b15}, since
this surface is the equilibrium of the gauge force. In addition,
fig. \ref{forcefield_gauge_2a15_2b15} illustrates that this
equilibrium is stable only for $x<1$, which justifies the definition
\eqref{GribovRegion} of the `Gribov region'.

\bfi
\centering
  \includegraphics[clip,width=0.92\columnwidth]{figs/forcefield_gauge_2a15_2b15.eps}
  \caption{The gauge fixing force of the toy model for $n=1$, $\alpha^{-1}=0.15$.}
\label{forcefield_gauge_2a15_2b15}
\efi

\bfi
\centering
\includegraphics[clip,width=0.92\columnwidth]{figs/xhisto_eta0d001_beta0_varsmalln.eps}
  \caption{`Entropy effect' for the (not gauge-invariant) distribution
    of values of $x$ at the parameters $\alpha=5\cdot 10^{-5}$, $\beta=0$, step size $\eta=0.001$.}
\label{xhisto_0alpha20000_3beta0_varlx}
\efi

\bfi
\centering
\includegraphics[clip,width=0.92\columnwidth]{figs/xhisto_eta0d0001_beta0_n500.eps}\vspace{0.4cm}
\includegraphics[clip,width=0.92\columnwidth]{figs/xhisto_eta0d0001_beta1400_n500.eps}\vspace{0.4cm}
\includegraphics[clip,width=0.92\columnwidth]{figs/xhisto_eta0d0001_beta1650_n500.eps}\vspace{0.4cm}
\includegraphics[clip,width=0.92\columnwidth]{figs/xhisto_eta0d0001_beta1800_n500.eps}
\caption{Increasing $\beta$, the strength of the gauge action, at $n=500$ for
$\alpha=10^{-4},\dots,10^{-8}$. Values of $\beta$ indicated in the plots. Step size $\eta=10^{-4}$.}
\label{f.toy3}
\efi

\subsection{Numerical simulation}\label{subsec:toynumerics}
To see explicitly the effects of the various parameters, a numerical
simulation of this toy model is easily possible, e.g.\ by means of a
random walk algorithm, in which the probability for a local update
with fixed step size $\eta$ (introducing $\{x_\nu\} ,\, \nu=0,\dots,n$
with $y\equiv x_0$) is given by
\begin{equation}
p(x_\nu \to x_\nu \pm\eta\hat{e}_\nu)=\frac{1}{2}\pm\frac{1}{2}\tanh
    \left(\frac{\eta}{2}K_\nu(\{x_\mu\})\right). \label{e.rwe}
\end{equation}
A gauge-invariant dynamics may be defined from the orbits
\begin{equation}
  u_i=x_i\exp\left(-\frac{x^2}{2}-y^2\right)(=\text{const.}),\ \ i=1,\dots,n
\end{equation}
as
\begin{equation}\label{action}
S=\beta u^2,\ \ u^2 = \sum_{i=1}^n u_i^2 .
\end{equation}
$S$ is monotonously increasing for $0 \leq x \leq 1$ and has therefore
the effect of driving the configurations toward $x=0$, similarly to
the expected effect of the QCD action.

The dynamical part of the force is given by \be
K_\nu^a=-\frac{dS}{du}\frac{du}{dx_\nu}, \ee
and the drift force of the random walk algorithm, eq. \eqref{e.rwe}, is
\be
K_\nu = K_\nu^a + K_\nu^g.
\ee
In the following figures we show the distribution of gauge-fixed
configurations, $\rho(x)$.  The increasing proximity of the
configurations to the Gribov horizon with larger $n$ is illustrated in
Fig. \ref{xhisto_0alpha20000_3beta0_varlx}. The effect of the gauge
fixing force is as expected, since the numerical results obey the
distribution \eqref{asymp_alpha_inf}. The most interesting
observations pertain the combined effect of the three parameters: $n,
\alpha$ and $\beta$. As illustrated in Fig. \ref{f.toy3}, keeping $n$
fixed, the action has little effect up to some value of $\beta$
(compare figs. \ref{f.toy3}a and \ref{f.toy3}b), and then suddenly
starts driving the configurations toward the interior of the Gribov
zone, selectively depending on $\alpha$, see figs. \ref{f.toy3}c and \ref{f.toy3}d. This
happens at smaller values of $\beta$ for lower $n$.\par
Notice that in all cases there is a small proportion of configurations
just outside the Gribov horizon. This is due to the fact that the
confinement to the Gribov region is achieved by return cycles
engendered by the unstable modes which $K^g$ develops beyond the
horizon, see fig. \ref{forcefield_gauge_2a15_2b15}. The fraction of
configurations outside the Gribov horizon shrinks as the strength of
the gauge fixing force increases, see fig. \ref{f.toy3}a.

The lesson from this toy model is then that there are subtle and
non-linear effects in the combined action of the gauge fixing, the
gauge dynamics and the entropy, and we should not be amazed to find
such non-trivial behaviour in the realistic case of lattice gauge
theory.

\section{Results for the gluon and ghost propagator}\label{sec:results}
\subsection{Technicalities}
The relation between the continuum and the lattice momenta is, as
usual,
\begin{equation}
  q_\mu(k_\mu)=\frac{2}{a(\beta)}\sin\left(\frac{\pi k_\mu}{L_\mu}\right).
\end{equation}
We impose the cylinder cut \cite{Leinweber:1998uu} on the momenta. The
propagators have been renormalised at $\mu=2.5\,\text{GeV}$. For some
details concerning the implementation of the ghost propagator, see
subsec. \ref{subsec:ghp}.
\par
We compare our results for the IR exponents with the `scaling
predictions' \cite{Lerche:2002ep}, which are $\kappa=0.2$ in $d=2$,
$\kappa\approx 0.4$ in $d=3$ and $\kappa\approx 0.6$ in $d=4$, with $\kappa_A$ determined by eq. \eqref{kAkCvarD}, see \ref{sec:irprops}.

\subsection{Gluon propagator}

We have calculated the scalar gluon propagator function in two, three
and four dimensions,
\begin{equation}
  D_\text{gl}(k)=\frac{1}{N_c^2-1}\frac{1}{d-1}\sum_{\mu,a}
  \left\langle A_\mu^a(k)A_\mu^a(-k)\right\rangle,
\end{equation}
with the standard lattice discretisation of the gauge field,
$A_\mu^a(x)\propto u_\mu^a(x)$. We find significant differences
between stochastic and standard gauge fixing at relatively large
$\Delta^2$. In turn, if we fix the gauge to a high precision,
$\Delta^2 \leq 10^{-15}$, the differences disappear, and we confirm
previous lattice results in all dimensions. Notice that at large
$\Delta^2$ the picture is complicated by the fact that the results of
standard STOR after a heat-bath thermalisation strongly depend on the
value of $p$, which fixes the probability of non-optimal gauge
transformations in the STOR algorithm. The differences mentioned are
directly related to the distance from $\partial\Omega$, which is
visible from the spectrum of the Faddeev-Popov operator, see section
\ref{sec:FPOS}.

\subsubsection{Two dimensions}\label{subsubsec:gluon-2d}
\bfi
\centering
\includegraphics[clip,width=0.92\columnwidth]{figs/glp_cylcut_ty23_ls200_2d.eps}
\caption{Gluon propagator on a $200^2$ lattice at
  $\beta=10$ ($a=0.082\text{fm}$).}
\label{glp_cylcut_ty23_ls200_2d}
\efi

The two-dimensional case is special; there are analytical results for,
e.g., the string tension \cite{Dosch:1978jt}, and the existence of
scaling behaviour has been viewed as relatively uncontroversial also
from lattice data \cite{Maas:2007uv,Cucchieri:2007rg}. Our results for
this case appear to corroborate the latter findings. The data are
shown in fig. \ref{glp_cylcut_ty23_ls200_2d}. We extract the infrared
exponent $\kappa_A=-1.37$, which corresponds to $\kappa=0.19$ if
eq. \eqref{kAkCvarD} is assumed to hold. This extraction is done via a
power law fit to the five most IR data points after discarding the two
lowest non-vanishing momenta, which closely resembles the method
employed in \cite{Maas:2007uv}. $\chi^2/\text{ndf}$ is safely below
1. As the value of $\kappa$ nicely agrees with the prediction of
$0.2$, this appears to speak in favour of the scaling solution in 2D
on the lattice, corroborating previous results obtained with standard
gauge fixing \cite{Maas:2007uv,Cucchieri:2007rg}. Even more
importantly, a comparison with the ghost propagator
(subsec. \ref{subsec:ghp}) shows that the scaling relation
\eqref{kAkCvarD} is approximately fulfilled. Note, however, the
surprising fact that the naive scaling hypothesis even fails in two
dimensions at $\beta=0$, see \cite{mpssvs}. Here it has been found that
while approximate scaling holds true for the gluon propagator for
intermediate momenta, the ghost propagator is subject to drastic
changes if changing the global properties of the gauge fixing.

\begin{figure}
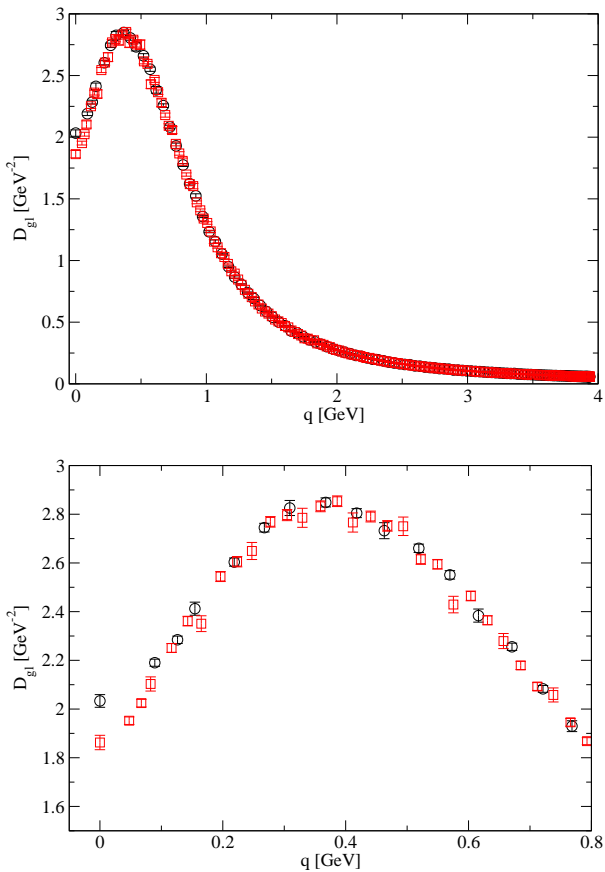

\centering
  \includegraphics[clip,width=0.92\columnwidth]{figs/glp_mr2d5GeV_beta4d2_varls_3d.eps}\vspace{0.4cm}
  \includegraphics[clip,width=0.92\columnwidth]{figs/glp_mr2d5GeV_beta4d2_varls_3d_zoom.eps}
\caption{\emph{Top (a):} Gluon propagator in 3D at $\beta=4.2$
  ($a=0.17\,\text{fm}$) on $80^3$ (black circles) and $150^3$ (red
  squares). \emph{Bottom (b):} Blowup of the same data in the IR.}
\label{glp_mr2d5GeV_beta4d2_varls_3d}
\end{figure}

\subsubsection{Three dimensions}

In three dimensions, it has been observed in lattice simulations
\cite{Cucchieri:2003di} that the gluon propagator peaks at
$q=350^{+100}_{-50}\,\text{MeV}$, but still approaches a non-vanishing
value in the infrared, $D(q^2\to 0)>0$. We confirm this observation
with stochastic gauge fixing on a $150^3$ lattice at $\beta=4.2$
($\hat{\approx}(26\,\text{fm})^2$), see
fig. \ref{glp_mr2d5GeV_beta4d2_varls_3d}. Furthermore, we observe a
sign of a finite volume effect when comparing the data to the $80^3$
case (blowup in fig. \ref{glp_mr2d5GeV_beta4d2_varls_3d}b). In
agreement with previous results for $d=3$ \cite{Cucchieri:2007md},
there is no evidence that this suffices to yield the scaling solution
as the volume $V\to\infty$.\par This is compatible with the findings
in \cite{mpssvs} for $\beta=0$. There it has been shown that the
scaling window within a standard gauge fixing moves towards larger
momenta. Naively translated to non-vanishing $\beta$, this suggests
that the assumed scaling region has to be washed out by the dynamics
if we go to higher dimensions.\par For comparison, we have also
performed simulations with standard gauge fixing on a $150^3$ lattice,
which yield a gluon propagator essentially identical with the one
obtained from stochastic quantisation.

\subsubsection{Four dimensions}\label{subsubsec:glp4d}
\begin{figure}
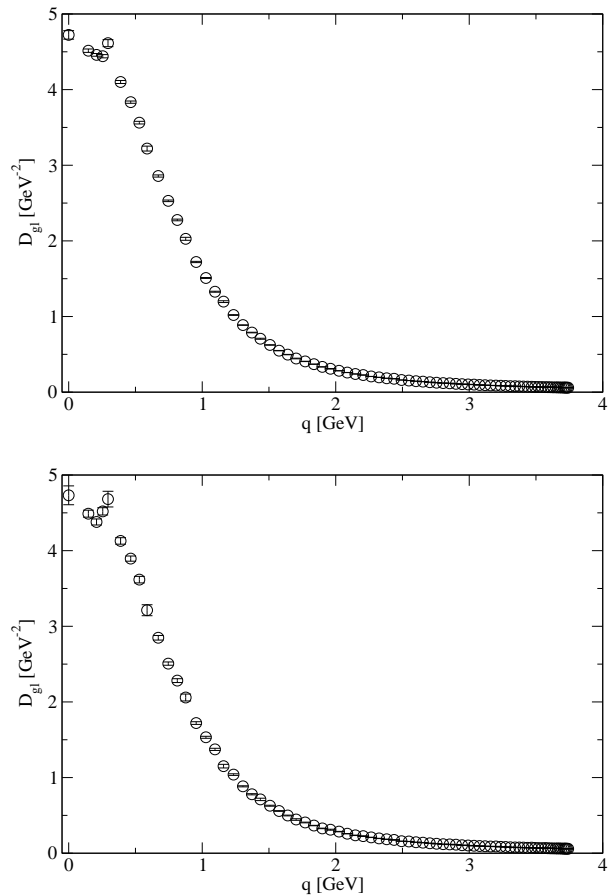

  \centering
 \includegraphics[clip,width=0.92\columnwidth]{figs/glp_cylcut_mr2d5GeV_beta2d2_ls40_4d.eps}\vspace{0.4cm}
 \includegraphics[clip,width=0.92\columnwidth]{figs/glp_cylcut_mr2d5GeV_ty26_beta2d2_ls40_4d.eps}
  \caption{\emph{Top (a):} Gluon propagator, $40^4$, $\beta=2.2$
    ($a=0.21\,\text{fm}$). \emph{Bottom (b):} For comparison, the result of
    a simulation with lower g.\,f. accuracy, sc.\ only
    $\Delta^2<10^{-8}$, but where also the `fine-tuning' is done by
    stochastic gauge fixing (see text).}
\label{glp_cylcut_mr2d5GeV_beta2d2_ls40_4d}
\end{figure}

For $d=4$, we again confirm earlier lattice results
\cite{Cucchieri:2007md,Bogolubsky:2007ud} for sufficiently accurate
stochastic gauge fixing. The gluon propagator on a $40^4$ lattice at
$\beta=2.2$ is, like in 3D, clearly finite in the IR.  Unlike in 3D,
it does not possess a peak at finite momentum, see
fig. \ref{glp_cylcut_mr2d5GeV_beta2d2_ls40_4d}. $V\approx(8.4\,\text{fm})^4$
is below the maximal lattice size already investigated with
standard methods e.g.\ in
\cite{Sternbeck:2007ug,Cucchieri:2007md}. But the lower-dimensional
results, including those on the spectrum of $-\partial D$,
sec. \ref{sec:FPOS}, suggest that the problem is not a finite volume
effect, but may rather be the well-known gauge fixing problem, as laid
out in sections \ref{sec:intro} and \ref{sec:irprops}.\par
Our value of the gauge functional $R=0.8278(1)$ essentially agrees
with the value obtained from standard gauge fixing, while it is
significantly below the value after improved gauge fixing aimed
specifically at maximising $R$ \cite{Bogolubsky:2007bw}. Here,
`standard gauge fixing' denotes a method of choosing first copies
after overrelaxation, and `improved g.\,f.' refers to employing
simulated annealing and $\mathds{Z}_2$ flips. This is consistent with
our other findings, indicating that the effect of stochastic gauge
fixing is similar to the one of standard g.\,f., which renders the
putative lattice gauge fixing problem more general.\par
Since our fine-tuning by stochastic overrelaxation may change the
location of the sampled Gribov copies, we have also produced some
results where the fine-tuning has been performed entirely within the
scheme of stochastic quantisation. In case of a random walk
implementation, this amounts to decreasing $\eta\to 0$ by some
prescription, e.g., exponentially. Even though the accuracy of gauge
fixing has been lessened here, the result is virtually the same,
see fig. \ref{glp_cylcut_mr2d5GeV_beta2d2_ls40_4d}b.

\bfi
\centering
\includegraphics[clip,width=0.92\columnwidth]{figs/ghdf_ls200_2d_forposter.eps}\vspace{0.4cm}
\includegraphics[clip,width=0.92\columnwidth]{figs/ghdf_ir_ls40et80_forposter_withfit.eps}\vspace{0.4cm}
\includegraphics[clip,width=0.92\columnwidth]{figs/ghdf_ls20et40_4d_forposter_withfit.eps}
\caption{Ghost dressing function in (a) $d=2$ \emph{(top)}, (b) $d=3$ \emph{(middle)} and
  (c) $d=4$ \emph{(bottom)}, at $\beta=2.2$, $4.2$ and $10$, respectively.}
\label{fig:ghdf234}
\efi

\subsection{Ghost propagator}
\label{subsec:ghp}
The ghost propagator on the lattice is given by
\begin{equation}
  D_\text{gh}(k)=\sum_{x,y}\left\langle
    \left(\mathds{M}^{-1}\right)^{ab}_{xy}\right\rangle e^{ik\cdot(x-y)},
\end{equation}
with $k\cdot x=2\pi\sum_\mu k_\mu x_\mu /L_\mu$. Its calculation is
computationally expensive, since it requires the inversion of the
lattice Faddeev-Popov operator $\mathds{M}$, whose standard form may
be found e.g.\ in \cite{Sternbeck:2008wg}. We have implemented both
the point source method, see e.g.\ \cite{Boucaud:2005gg}, and the
plane wave source method, see e.g.\ \cite{Cucchieri:1997dx}. This
means that $\mathds{M}$ is inverted either on a point source
$s_b^{a,x}=\delta^{ab}(\delta_{x,0}-1/V)$ or a vector of plane waves
$s_b^{a,x}(k)=\delta^{ab}e^{ik\cdot x}$. While the former method leads
to heavy fluctuations in the UV, the latter one is much more
time-consuming. The results shown here have been produced with a plane
wave source, but sometimes inverting the Faddeev-Popov operator only
on a proper subset of the momenta surviving the cylinder cut.\par
In two dimensions ($d=2$), we extract from the ghost propagator, like
from the gluon propagator, an infrared exponent close to the expected
value $\kappa=0.2$, sc.\ $\kappa=0.17$. The basic fitting method is the
same as for the gluon data, see \ref{subsubsec:gluon-2d}. See
fig. \ref{fig:ghdf234}a for results on a rather large lattice.\par

For $d=3$, we expect to see a finite-volume effect for the ghost
  propagator for the volumes used here, $40^3$ and $80^3$. A fit of the
  $40^3$ data in fig. \ref{fig:ghdf234} to a power law yields
  $\kappa=0.25$, which is already clearly below $\approx 0.4$, i.e.,
  the `scaling prediction'. A fit of the $80^3$ data in
  fig. \ref{fig:ghdf234} to a power law yields $\kappa\approx 0.2$,
  indicating clear finite volume effects.\par
For $d=4$, a similar effect is apparent from
fig. \ref{fig:ghdf234}c. On a small lattice, sc.\ $20^4$, we obtain
$\kappa=0.26$. Again, this is far below the `scaling prediction' of
$\approx 0.6$. The discrepancy is significantly larger than for
$d=3$. Moreover, the scaling relation \eqref{kAkC4D} is violated
thereby, given the gluon result of $\kappa_A=-1$
(subsubsec. \ref{subsubsec:glp4d}), which implies $\kappa=0.5$ under
\eqref{kAkC4D}. On the larger lattice, the IR behaviour is again
harder to fit precisely, but clearly less divergent. Here, we have
produced additional data with the point source method. This introduces
heavy fluctuations in the UV, but does less harm in the IR, which is
our focus of interest here. From these data, we obtain
$\kappa=0.19$. -- The fact that $\kappa$ doesn't vanish may well be
due to the finite lattice extension
\cite{Bogolubsky:2009dc,Cucchieri:2008fc} or also to choosing the
first instead of the `best' Gribov copy \cite{Bornyakov:2008yx}.

\section{Spectrum of the Faddeev-Popov operator}\label{sec:FPOS}
\subsection{Motivation and technicalities}
As the first Gribov region $\Omega$ is the set of configurations with
positive Faddeev-Popov operator (FPO), $-\partial D>0$, the lowest
eigenvalue of $-\partial D$ vanishes at the Gribov horizon
$\partial\Omega$. Hence, its value may be interpreted as a measure of
the distance from $\partial\Omega$. This is an interesting test of the
Gribov-Zwanziger scenario of confinement, since this predicts that
configurations near $\partial\Omega$ account for confinement. The
lattice discretisation of the FPO is a $3V\times 3V$ real symmetric
matrix. Previous lattice studies have established that the gauge
copies tend to move towards the Gribov horizon as $V$ is increased,
and away from it as the gauge fixing is improved, i.e., as $R$ is
increased \cite{Sternbeck:2005vs,Sternbeck:2005re}.\par
We want to compare the spectra after stochastic and standard gauge
fixing. The fact that we intend to do so also at insufficient gauge
fixing (i.e., relatively large $\Delta^2$) imposes a restriction on
the method used to extract the eigenvalues of the FPO. In particular,
we cannot assume that the Faddeev-Popov matrix is positive
definite. This prevents us from applying the conjugate gradient method for the matrix inversion and exploiting its connection to 
the Lanczos algorithm for finding a subset of its eigenvalues
\cite{MeurantStrakos06}. Thus, we find \emph{all} eigenvalues
of the $3V\times 3V$ Faddeev-Popov matrix with a standard algorithm
for real symmetric matrices, sc.\ symmetric bidiagonalisation and QR
reduction, which is rather impracticable on large lattices.\par
The standard lattice discretisation of the FPO which we use is the
Hessian of the gauge fixing functional \eqref{stdgff}. As usual, terms
that vanish under the gauge fixing condition are omitted.  To employ
this discretisation also at imperfect gauge fixing is clearly not a
unique choice, but has the advantage that the three zero modes of the
FPO exist to a better approximation. The `lowest nontrivial FPO
eigenvalue $\lambda_0$' is always understood without these three
trivial eigenvalues. While this choice requires inferences from the
value of $\lambda_0$ to the distance to the Gribov horizon to be taken
with a grain of salt, it still permits a qualitative comparison
between stochastic and standard gauge fixing.

\subsection{Results}\label{subsec:FPOS_results}
\begin{figure*}
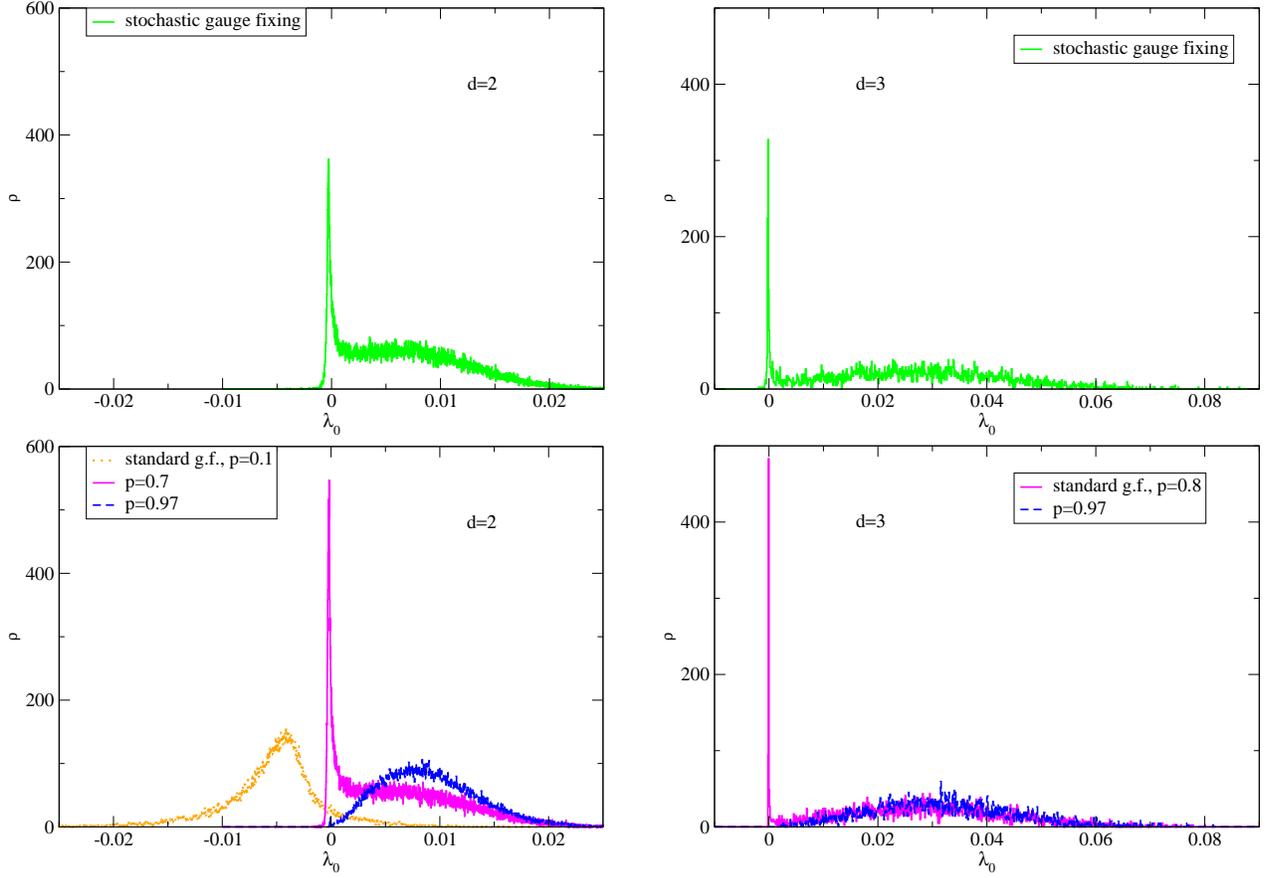

\centering
\mbox{
\includegraphics[clip,width=0.92\columnwidth]
{figs/FPO_merged_ls24_2d_ity19vs25varp-b.eps}
\qquad
\includegraphics[clip,width=0.92\columnwidth]
{figs/FPO_merged_ls12_3d_ity19vs25varp-b.eps}
}
\mbox{
\includegraphics[clip,width=0.92\columnwidth]
{figs/FPO_merged_ls24_2d_ity19vs25varp-a.eps}
\qquad
\includegraphics[clip,width=0.92\columnwidth]
{figs/FPO_merged_ls12_3d_ity19vs25varp-a.eps}
}
\caption{Histograms of the lowest nontrivial FPO eigenvalue $\lambda_0$ in
  $d=2$ resp. $d=3$ from stochastic \emph{(top)} and standard gauge fixing \emph{(bottom)}. \emph{Left:} $24^2$, $\beta=10$, $\Delta^2\approx
  8\cdot 10^{-4}$. \emph{Right:} $12^3$, $\beta=4.2$, $\Delta^2\approx
  7\cdot 10^{-4}$.}
\label{fig:FPO_merged_2d3d}

\end{figure*}

\bfi
\centering
\includegraphics[clip,width=0.92\columnwidth]{figs/glp_mr2d5GeV_ls24_2d_ty23_varp.eps}
\caption{Gluon propagators corresponding to the lower left plot of
  fig. \ref{fig:FPO_merged_2d3d}.}
\label{glp_mr2d5GeV_ls24_2d_ty23_varp}
\efi
\bfi
\centering
\includegraphics[clip,width=0.92\columnwidth]{figs/scatter_lambda0corrD0_varp.eps}
\caption{Scatter plot of lowest FPO eigenvalue vs. gluon propagator at $q=0$ for intermediate $\Delta^2$ (data like in
  lower left plot of fig. \ref{fig:FPO_merged_2d3d}). Correlation
  coefficient: $\rho=-0.40$ ($p=0.1$, $N=36000$ configurations)
  resp. $\rho=-0.29$ ($p=0.97$, $N=48000$). For clarity, only $10\%$ of data points are shown.}
\label{scatter_lambda0corrD0_varp}
\efi

\bfi
\centering
\includegraphics[clip,width=0.92\columnwidth]{figs/FPO_smallDelta_stoch.eps}\vspace{0.4cm}
\includegraphics[clip,width=0.92\columnwidth]{figs/FPO_smallDelta_std.eps}
\caption{Histograms of $\lambda_0$ on a $24^2$ lattice at $\beta=10$. \emph{Top (a):} Stochastic gauge fixing with increasing accuracy by choosing a small step size during the entire stochastic process ($\Delta^2\approx 1.5\cdot 10^{-6}$) resp. decreasing the step size for fine-tuning ($\Delta^2<10^{-10}$ and $\Delta^2<10^{-15}$). \emph{Bottom (b):} Standard gauge fixing with $\Delta^2<10^{-10}$ for different values of $p$.}
\label{FPO_smallDelta}
\efi

\begin{figure}
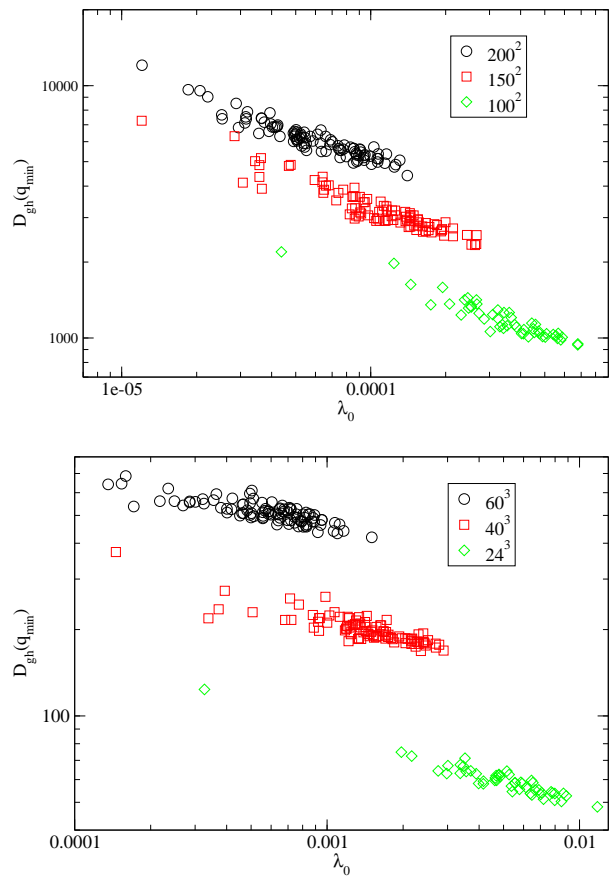

\centering
  \includegraphics[clip,width=0.92\columnwidth]{figs/eval0_corr_ghp0_varls_2d.eps}\vspace{0.4cm}
  \includegraphics[clip,width=0.92\columnwidth]{figs/loweval_corr_ghp0_bc0_ls24et40et60_3d_forpaper.eps}
  \caption{Scatter plot of ghost propagator at lowest non-vanishing
    momentum vs. approximate lowest nontrivial FPO
    eigenvalue. \emph{Top (a):} $d=2$. Correlation coefficient $r=-0.82$
    resp. $-0.83$ in all three cases. \emph{Bottom (b):} $d=3$. $r=-0.74$ resp. $-0.79$ in all three cases. Between $50$ and $100$ configurations per lattice size.}
\label{loweval_corr_ghp0_2d3d}
\end{figure}

A few results on the Faddeev-Popov eigenvalues have already been
presented in the analysis of return cycles,
subsec. \ref{subsec:retcyc}; the bulk of results is shown in the
following.\par
While we have investigated the full Faddeev-Popov operator spectrum
mainly on tiny lattices, like $24^2$, the picture we have obtained is
rather unambiguous. Let us first state the results for stochastic
gauge fixing, i.e., random walk steps with `locally' included gauge
fixing. At some intermediate $\Delta^2$, i.e., at a not very small
distance from $\Gamma$, the distribution of the lowest non-vanishing
eigenvalue of the Faddeev-Popov operator after stochastic gauge fixing
shows a very pronounced peak close to the Gribov horizon. The peak is
located slightly outside the Gribov horizon, see
fig. \ref{fig:FPO_merged_2d3d} (upper plots). This implies that Landau gauge is not
fixed properly at this $\Delta^2$, since $-\partial D$ is the Hessian of $\int {d^4x
  |^gA|^2}$, which is minimised by numerical Landau gauge fixing. \par

In the case of standard gauge fixing, i.e., STOR gauge fixing
after heat-bath thermalisation, the distribution of $\lambda_0$ exhibits a twofold dependence on $\Delta^2$ and the STOR parameter $p$. It resembles its counterpart from stochastic gauge fixing only around some specific value of $p$. Let us spell this out in more detail. At intermediate $\Delta^2$ and large $p$, the
procedure yields a peak at some distance from the horizon and clearly
inside of it. When the accuracy of the standard gauge fixing is further 
lowered by stopping STOR at a larger $\Delta^2$, the distribution is shifted outside of the horizon, without a peak especially close to $\partial\Omega$. At intermediate $\Delta^2$, we compare different values of $p$. At small $p$, much more
configurations are outside the horizon, and at intermediate $p$, the
observation from stochastic gauge fixing is qualitatively
reproduced, see fig. \ref{fig:FPO_merged_2d3d} (lower plots). Interestingly, the latter value roughly minimises the
number of gauge fixing sweeps to obtain the desired value of $\Delta^2$. For sufficiently small $\Delta^2$, the distribution obtained from stochastic gauge fixing coincides with the one from standard gauge fixing regardless of $p$, and no bias towards the Gribov horizon remains. This is evident from the contrast between the left plots of fig. \ref{fig:FPO_merged_2d3d} and their counterparts at smaller $\Delta^2$, fig. \ref{FPO_smallDelta}.\par

Moreover, fig. \ref{glp_mr2d5GeV_ls24_2d_ty23_varp} illustrates that
at the same intermediate value of $\Delta^2$, the gluon propagator is less similar
to the fully gauge-fixed result if more configurations are outside the
Gribov region ($\lambda_0<0$). This difference is created, as
described, by varying $p$. The gluon propagator at vanishing momentum
is significantly anticorrelated with $\lambda_0$, see
fig. \ref{scatter_lambda0corrD0_varp}.\par

A peak close to $\partial\Omega$, but of course inside $\Omega$ is of interest
especially for the Gribov-Zwanziger scenario of confinement, since it would imply that those configurations are preferably sampled which account for confinement according to this scenario. However, generating
configurations at smaller $\Delta^2$ by adjusting $\alpha$ and the
step size does not help to generate such a peak: For small $\Delta^2$, the distributions from
stochastic and standard gauge fixing become indistinguishable, regardless of $p$. This
is also true if $\Delta^2$ is lowered gradually by decreasing $\alpha$
and $\veps$ resp. $\eta$ simultaneously.\par

It is possible to relate the distribution after stochastic gauge
fixing at intermediate $\Delta^2$ to a speculative scenario involving `return
cycles', see subsec. \ref{subsec:retcyc}. In this intuitive picture,
the Langevin evolution leads to an accumulation of the gauge-fixed
configurations at the Gribov horizon.\par

Moreover, we find a strong anticorrelation between the IR ghost
propagator and the lowest FPO eigenvalue, confirming findings of
\cite{Sternbeck:2005re,Sternbeck:2005vs} with standard gauge
fixing; see fig. \ref{loweval_corr_ghp0_2d3d} (unrenormalised
data). Note that on the larger lattices, we have not calculated all
FPO eigenvalues, but rather employed the method of
\cite{MeurantStrakos06}. Thus, in fig. \ref{loweval_corr_ghp0_2d3d},
`$\lambda_0$' does not refers to the strictly lowest eigenvalue, but
to the smallest one as found by the Lanczos procedure related to the
conjugate gradient method.\par
Finally, we also observe a finite volume effect, sc.\ that the lowest
eigenvalue tends to be smaller on larger lattices, again like already
reported in \cite{Sternbeck:2005re,Sternbeck:2005vs}.

\section{Conclusion}

In the present work we have studied lattice Yang-Mills theory in
Landau gauge within stochastic quantisation in two, three and four
dimensions.  We also have studied a finite-dimensional toy model in
order to illustrate some subtleties of gauge-fixed Langevin
evolutions. In this model we see an interplay between the
dimensionality of the system, the gauge force and the dynamics induced by the action.

Our results for the ghost and gluon propagator and the
low-lying eigenvalues of the Faddeev-Popov operator agree with the
lattice results obtained so far with standard gauge fixing procedures.
The detailed analysis revealed an interesting approach towards the
final distribution on the first Gribov region if the gauge fixing is
bettered. For worse gauge fixing quality the distribution shows also a sharp peak
slightly outside the Gribov horizon. This implies a statistical bias
towards the Gribov-Zwanziger scenario. We conclude that stochastic
quantisation provides a tool of how to avoid a possible mechanism
against the Gribov-Zwanziger scenario. Still this does not suffice to
adjust the gauge fixing appropriately, i.e., to obtain on the
lattice the full one-parameter family of solutions found in continuum
studies. This could suggest that the gauge fixing problem on the
lattice might be even more general than previously assumed.

Together with further evidence from $\beta=0$ computation on the
lattice \cite{mpssvs}, as well as recent work on Landau gauge fixings
\cite{Maas:2009se}, our results hint at the severity of the Gribov
problem in lattice simulations. We hope that this problem can be
resolved in near future along the ways outlined in
\cite{Maas:2009se,mpssvs}, and in particular in
\cite{vonSmekal:2007ns,vonSmekal:2008es}.

\acknowledgments 
We thank M.~Ilgenfritz, A.~Maas, A.~Nakamura, L.~von Smekal and D.~Zwanziger for
discussions. This work is supported by Helmholtz Alliance
HA216/EMMI. D.S. acknowledges support by the Landesgraduiertenf\"orderung Baden-W\"urttemberg via the Research Training Group ``Simulational Methods in Physics''. I.-O.S. is indebted to the MPI for Physics, M\"unchen for hospitality. The numerical simulations were carried out on the compute
cluster of the ITP, University of Heidelberg, and on bwGRiD
(http://www.bw-grid.de), member of the German D-Grid initiative.

\bibliography{bib/intro}
\bibliographystyle{apsrev-SLAC}
\end{document}